\documentclass[a4paper]{jpconf}
\usepackage{graphicx}
\usepackage{amsmath, amssymb}

\begin{document}
\title{Lightning in other planets}

\author{Christiane Helling}

\address{Centre for Exoplanet Science,  University of St Andrews, St Andrews, KY16 9SS, UK\newline
SUPA, School of Physics \& Astronomy, University of St Andrews, St Andrews, KY16 9SS, UK\newline
SRON Netherlands Institute for Space Research, Sorbonnelaan 2, 3584 CA Utrecht, NL}

\ead{ch80@st-and.ac.uk}

\begin{abstract}
More than 4000 planet are known that orbit stars other than our
Sun. Many harbor a dynamic atmosphere that is cold enough that cloud
particles can form in abundance. The diversity of exoplanets leads to
differences in cloud coverage depending on global system
parameters. Some planets will be fully covered in clouds, some have
clouds on the nightside but are largely cloud-free on the dayside. These
cloud particles can easily be charged and lightning discharges will
occur in cloudy, dynamic exoplanet atmosphere.  Lightning supports a
Global Electric Circuit (GCE) on Earth and we argue that exoplanet may
develop a GCE in particular if parts of the exoplanet atmospheres can remain cloud
free.
\end{abstract}

\section{Introduction}
Today, about 4070 planets are known that orbit stars other than
our Sun. No Earth-twin has been found so far, instead, planets bigger
than Earth (super-Earths) or bigger than Jupiter (giant gas planets),
smaller than Neptune (mini-Neptune), to name a few, were
discovered. Several of the known exoplanets orbit their host star very
closely such that a rocky surface turns into magma or Jupiter-like
exoplanets expand their atmosphere considerably. Inside the
atmospheres of exoplanets, clouds form and the cloud particles (also
called {\it aerosols}) are made of a mix of minerals, for example
silicates and iron oxide
\cite{annurev-earth-053018-060401}.
Spectroscopic
studies of exoplanets are repeatedly frustrated by clouds blocking the
view into the underlying atmosphere to reveal the existence of
potential biosignatures. Observations of the planet's thermal emission
show that the atmosphere of giant gas planets may have very dynamic
global circulation \cite{2014Sci...346..838S} due to a strong
day/night temperature difference. Although the solar system gas
planets have far less dynamic atmospheres globally, lightning events
are directly detected for the cloudy atmospheres of Earth, Jupiter,
and Saturn, are debated for Venus, and indirectly inferred for Neptune
and Uranus in our solar system. Sprites and high-energy particles are
observed above thunderclouds on Earth
\cite{2013SGeo...34....1F,2013ERL.....8c5027F}, and are theoretically
predicted for Jupiter, Saturn, and Venus.  On Earth, lightning
measurements can be conducted in situ, but the lightning statistics
for all other solar system planets are rather incomplete
\cite{2016MNRAS.461.3927H,2018PEPS....5...34L}. Other possible analogs for lightning on
exoplanets are terrestrial volcanoes that produce lightning during an
eruption inside the plume \cite{2000JGR...10516641J}. Studying
lightning on other planets inside and outside our solar system is of
interest because it enables us to study the electrostatic character of
such alien atmospheres, but also opens new possibilities for tracking
the dynamic behavior and the associated chemical changes in such
extraterrestrial environments.  The presence of an atmosphere that
shelters life may enable the formation of electrified clouds and the
emergence of lightning, and hence, the occurrence of an extrasolar
Global Electric Circuit (eGEC). In fact, lightning discharges are also
studied in other astronomical objects, in brown dwarfs and
planet-forming disks
\cite{2016PPCF...58g4003H,2000Icar..143...87D}. The fascination of
this research field lays in the universality of the underlying
physical processes which would benefit from a multilateral exchange of
ideas \cite{2016SGeo...37..705H} in contrast to siloing approaches.

\section{Lightning in other planets inside and outside our solar system}
\subsection{How much charges are required for lightning in exoplanet clouds}
Lightning occurs if an ensemble of cloud particles has acquired enough
charges and if a sufficient electrostatic potential is established to
overcome the local breakdown field of the atmospheric gas. A number of
free electrons is needed to build an unstable ionisation front that
eventually develops into the lightning plasma channel.  For
astrophysical purposes, the critical electric breakdown field is only
marginally depending on the chemical composition of the gas (unless
the gas is made of compounds of heavy elements only). The values for
the minimum electric potential, $V_{\rm t, min}$, for gases (mixes of
CO$_2$, N$_2$, O$_2$, H$_2$, He, CH$_4$) that dominate the solar
system planets varies between $\approx 400\,\ldots\,620$V (Table 1 in
\cite{2013ApJ...767..136H}).  The solar system's gas planets Saturn,
Jupiter, Uranus and Neptune have atmospheres dominated by
H$_2$/He/CH$_4$ (in this order) which is a representative starting
point for extrasolar planets, too. The more challenging question is
how many charges an ensemble of mineral cloud particles carries under
certain environmental conditions. This question can be approached
experimentally \cite{2014RScI...85j3903B,2018ApJ...867..123M} or
theoretically. It is tied to the cloud formation processes which
determine the composition, sizes and size distribution of the cloud
particles. The cloud formation process in exoplanet atmospheres is
strongly affected by changing thermodynamic conditions inside the
atmosphere. This results in cloud particles changing their sizes and
their material composition (e.g. Fig.~3 in
\cite{2018arXiv181203793H}): Small particles ($\approx
10^{-7}\,\ldots\,10^{-5}$cm) made of a mix of silicates and oxides
(e.g. MgSiO$_3$[s], Mg$_2$SiO$_4$/Fe$_2$SiO$_4$[s], SiO[s],
SiO$_2$[s], MgO[s], Fe$_2$O$_3$[s]; [s] stands for 'solid') compose
the upper and middle part of the clouds (T$_{\rm
  gas}<1000\,\ldots\,$1500K, $p_{\rm gas}<10^{-5}$bar ) and can grow
to $10^{-4}\,\ldots\,1$cm in deeper regions with $p_{\rm gas}>1$ bar
where the gas temperatures reach 1800K
\cite{2016A&A...594A..48L}. These big cloud particles are made of
mixes of high-temperature condensates (Fe[s], CaTiO$_3$[s],
Al$_2$O$_3$[s], TiO$_2$[s]).  The gas phase is collisional dominated
in most of these atmosphere regions, but the cloud particles
frictionally decouple quickly, hence, they fall into the atmosphere
with a velocity different to that of the gas. The energy balance
between frictional and latent heat, and collisional and radiative
cooling results in a $\Delta T\approx 3$K only
\cite{2003A&A...399..297W}.  The relative velocity between gas and
cloud particles is, however, insufficient to charge the exoplanet
mineral cloud particles. Turbulence driven particle-particle collisions that
result from  larger-scale convective-hydrodynamic motions  provide
the required kinetic energy to overcome the work function of the
materials that the cloud particles are made of
\cite{1980A&A....85..316V,2011ApJ...737...38H}. Charges can only
effectively separate between two particles by collisions
\cite{2008SSRv..137..335S}, hence, liquid droplets do not charge
easily.

The critical charge density, $\sigma_{\rm crit}$, or the number of
charges per cloud particle (Eq.~\ref{eq:apart}) or per total cloud
surface per volume (Eq.~\ref{eq:atot}) can be estimated for a given
minimal breakdown voltage, $V_{\rm t, min}$, and a geometrical
distance of $d$ of two charge-carrying surfaces,
\begin{eqnarray}
\sigma_{\rm crit} &=& \frac{\epsilon_0\,d\,V_{\rm t, min}}{r^2} \\
&\Rightarrow & \sigma_{\rm crib} \times A^{\rm tot}_{\rm cloud}  \,\,\,\,\,\,[e^- \mbox{cm}^{-3}]
\label{eq:atot}\\
&\Rightarrow & \sigma_{\rm crib} \times \frac{ A^{\rm tot}_{\rm cloud} }{n_{\rm d}} \,\,\,\,\,\,[e^- \mbox{grain}^{-1}].
\label{eq:apart}
\end{eqnarray}
The charge distribution is assumed to be spherical and of radius $r$
(being either the cloud particle size or the cloud extension;
$\epsilon_0= 8.85\,10^{12}$ F m$^{-14}$ -- vacuum dielectric constant
or electric permittivity of free space). $n_{\rm d}$ and $A^{\rm
  tot}_{\rm cloud}$ are the local number density of cloud particles
and the total cloud particle surface per unit volume of atmospheric
gas, respectively. The critical charge density per particle needed for an electric
field breakdown to initiate an ionisation front as precursor for a
lightning strike ranges from $10^{10}$ $e^- \mbox{grain}^{-1}$ in
low-pressure regions and and $10^{5}$ $e^- \mbox{grain}^{-1}$ at
higher pressures in an exoplanet atmosphere. The exact values will
vary for different atmospheric conditions in different extrasolar
planets.  The number do, however, allow us to make some first
comparison with laboratory experiments. Laboratory experiments are an
essential tool to test our basic assumptions as modelling the
individual charge processes of cloud particles that change size and
composition remains a challenge.  But also experiments come with
challenges as they mainly represent Earth-like conditions with respect
to thermodynamics and/or particle composition; a short summary is
given in \cite{2013ApJ...767..136H}. Experiments on frictional
electrification of KCl[s] and ZnS[s] particles
\cite{2018ApJ...867..123M} demonstrate that they attain charge
densities similar to those found on volcanic ash particles of
$10^{5}\,\ldots\,10^{8}$ $e^- \mbox{grain}^{-1}$. Both are consistent
with our theoretical estimates for exoplanet clouds.

\subsection{How lightning leaves chemical tracers}
Lightning introduces a drastic change of the atmospheric thermodynamic
properties locally.  Only during the first stage of the discharge
process is the gas chemistry driven by kinetic processes
\cite{2013JGRA..118.5190P} as result of the fast electrons impacting
the gas phase. The chemical composition is determined by the local
temperature at later stages when thermodynamic equilibrium has been
reached. In thermodynamic equilibrium, the drastic increase in
temperature alone cause the concentration of molecules that are
generally abundant in oxygen-rich planetary atmospheres of solar
element composition to drop by orders of magnitudes
\cite{2014ApJ...784...43B}. This includes molecules involved in the
formation of cloud particles like H$_2$O, SiO, MgOH, FeO, and others
like TiO, CO, CH$_4$. The concentration of CH increases and larger
carbohydrate molecule show a marginal increase. While lightning will
locally affect the gas-phase chemistry, it is reasonable to expect
that there will be no substantial feedback on the cloud formation
processes other than due to creating a bulk of free electrons that can
attach themselves to the surfaces of the cloud particles.

HCN is suggested to be a byproduct of lightning in the solar system
giant planets \cite{1980Sci...210.1351L}, or any discharge for that
matter, that could exist long enough to potentially be observable also
in extrasolar planets. HCN would be present for 2-3 yrs in the
detectable, optical thin atmosphere at $\lesssim 0.1$ bar after the
lightning event in the atmosphere of the exoplanet HAT-P-11b (R$_{\rm
  P}\approx 5$R$_{\rm Earth}$, M$_{\rm P}\approx 26$M$_{\rm Earth}$)
\cite{2016MNRAS.461.1222H}. For a mJy- radio emission to be detectable
at 150 MHz for HAT-P-11b, a flash rate $\sim4\times 10^5$ flashes km
$^{-2}$ h$^{-1}$ is required.

\section{The extrasolar Global Electric Circuit (eGEC)}
The presence of an atmosphere that shelters life, the formation of
electrified clouds and large-scale discharges under the affect of an
external radiation field lead to the emergence of a Global Electric
Circuit (GEC) on Earth which impacts the daily weather and traces the
global climate (\cite{2009AtmRe..91..140W}). A global electric
circuit is an electric current between charge-separating (i.e. clouds)
and non-charge-separating (i.e. cloud-free, 'fair weather') regions in
an atmospheres. The atmospheric gas in cloud-free regions has a higher
conductivity than in cloudy regions, hence, a leaking current
establishes, which was first discovered for Earth
\cite{wilson1906}. Can global electric circuits occur on extrasolar
planets?  A global electric circuit requires an upper and a lower
conducting region, charge separating processes and current flows
\cite{2016SGeo...37..705H,2012SSRv..168..363R,2008pae..book...11A}. Lightning
plays a key role in discharging the cloudy regions and in closing the
electric current cycle by driving a return current into the upper
atmosphere, the ionosphere on Earth. Without lightning, the
atmosphere's leaking current would equilibrate atmospheric potential
differences quickly.

 Three aspects need consideration in view of the large diversity of
 the known exoplanets and regarding the possibility that we may not
 yet be aware of all the physio-chemical processes involved. We
 explicitly include brown dwarfs into this discussion because they
 have atmospheric thermodynamic conditions comparable to giant gas
 planets and also form clouds in their atmospheres. {\it Firstly},
 only exoplanets that have an atmosphere will form clouds with
 atmospheric lightning that creates the required return current. This
 excludes exoplanets comparable to Mercury in size, proximity to the
 host star, planetary mass and radius. According to our present
 knowledge, brown dwarfs always have an atmosphere. While lightning
 (as discharges in general) will occur in other electrified
 environments (like volcanoes on Earth, dust devils on Mars), we will
 confine ourselves to atmosphere environment to be able to benefit
 from knowledge about the presently known ensemble of exoplanets (and
 brown dwarfs). {\it Secondly}, we discuss the formation of an
 ionosphere as upper conducting regions and argue for a lower
 conducting region in exoplanets and brown dwarfs. {\it Thirdly}, we
 need to consider where in the exoplanet atmosphere clouds may form as
 this has implication for the emergence of an eGEC.

 \begin{figure}
\includegraphics[width=15.5cm]{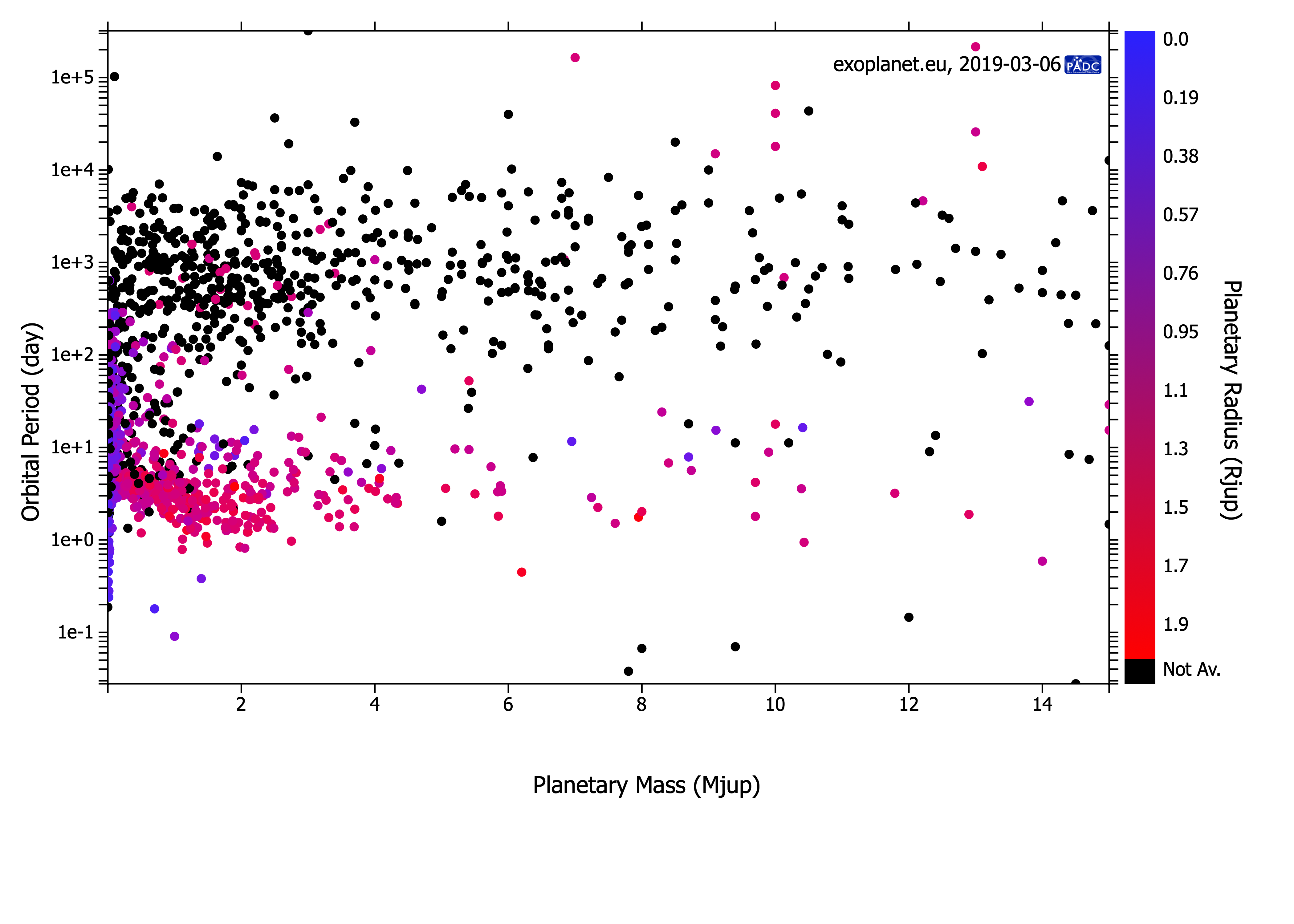}\\*[-2.7cm]
\includegraphics[width=15cm]{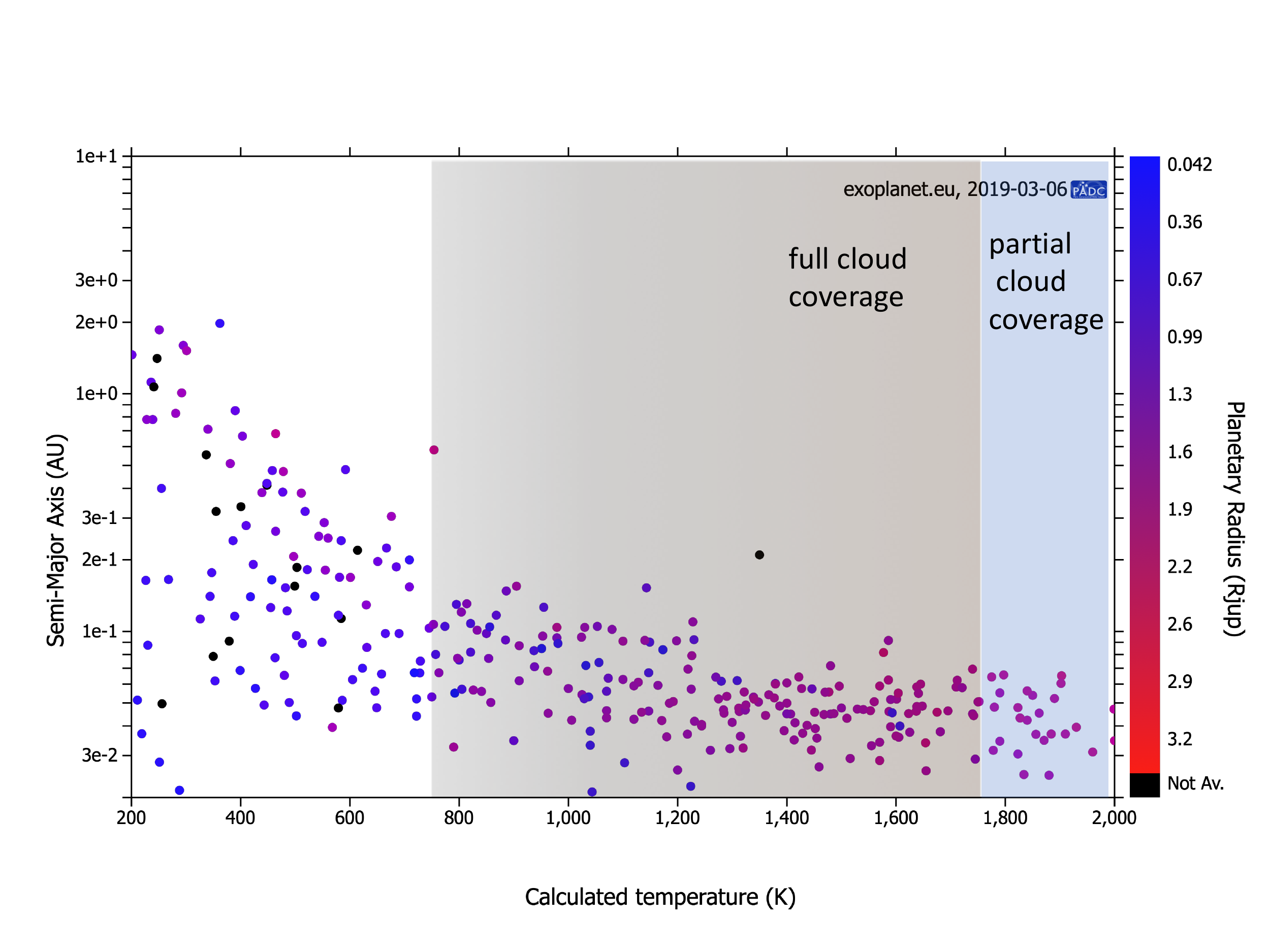}\\*[-0.8cm]
\caption{Extrasolar planets differ largely with respect to their
  orbital distance, masses and radii (top) which affect the global
  ('calculated') temperature (bottom). The
  combination of orbital distance, planetary mass, radii and global
  temperature provide a first indication for the presence of clouds,
  the possibility of lightning, and if a extrasolar global electric
  circuit can develop. The shaded areas are regimes of {\it full cloud
    coverage} (gray shaded) and {\it particle cloud coverage} (blue
  shaded). Lightning can occur in cloud-forming exoplanets, but a
  global electric circuit may establish in planets with partial cloud
  coverage only.  Note: M$_{\rm Earth}\approx3\cdot10^{-3}$ M$_{\rm
    Jup}$, R$_{\rm Earth}\approx9\cdot10^{-2}$, R$_{\rm Jup}$ T$^{\rm
    global}_{\rm Earth}\approx250$K, T$^{\rm global}_{\rm
    Jup}\approx150$K. [plots based on exoplanet.eu]}
\label{diversity}
 \end{figure}
 
 \paragraph{Upper conducting regions: Ionisation of the upper atmosphere: }
The presence of an upper conducting region is determined by the
irradiation (incl. cosmic rays) that a planet or a brown dwarf
receives.  This will depend on the host star (hot vs cool, active
vs. inactive). For a given host star, the flux impacting the upper
atmosphere will depend on the orbital distance, $a$, as $1/a^2$
between planet and star. Brown dwarfs occur as single stars in the
interstellar medium and have recently been found in binary systems
with white dwarfs. Both, the interstellar medium and the white dwarfs
will have a larger flux of high-energy radiation than most of the
presently known planet-hosting stars.  The major difference for the
formation of an ionosphere is therefore the ionisation process of the
upper atmospheric layers. For exoplanets orbiting close to their host
star, the very uppermost layers of a planetary atmosphere will be
ionised by stellar photons predominantly. Their intermediate
atmosphere layers will be heated by the thermal effect of photons
being absorped by the atmospheric gas which leads to a gas temperature
of 3500K on the dayside of super-hot Jupiter like WASP-18b or HAT-P-7b
\cite{2018A&A...617A.110P,2019arXiv190108640H}. No clouds will form
here. The nightside can be 2500K colder at similar pressures and
clouds can therefore form. In brown dwarfs and planets with large
semi-major axis (Fig.~\ref{diversity}), high-energy radiation fluxes
may create a shell of almost completely ionised gas in the outer
atmosphere \cite{2018A&A...618A.107R}. It is therefore reasonable to
assume that, similar to Earth, an electron current forms from these
layers of ionised gas to the boundary layers of the clouds.  The free
electrons attach themselves to the surface of cloud particles which,
hence, makes the gaseous charges particular immobile compared to cloud-free
parts of atmospheres. The conductivity of the remaining gas phase is
therefore low before the discharge, and hence, supports the
electrostatic potential difference.

\paragraph{Lower conducting regions:}
Exoplanets come in many sizes and temperatures (Fig.\ref{diversity}),
they can be rocky or have large gaseous envelope with only a small
rocky core. Brown dwarfs are heavier and form as stars, hence, they
are gaseous throughout. The lower conducting region is therefore,
similar to Earth, the rocky surface of a rocky planet. Depending on
its place in its planetary system, this surface maybe lava-like. Giant
gas planets have warmer or even hot inner atmosphere temperatures of
2000K$\,\ldots\,$3500K.  Warmer brown dwarfs reach temperatures of
upto 4000K in their inner atmosphere. We therefore suggest that this
thermally ionised inner atmosphere part of giant gas planets and brown
dwarfs serves as lower, inner conducting region because thermal
ionisation creates a partially ionised gas (e.g.,  Fig.~2 in
\cite{2015MNRAS.454.3977R}).  Ohmic heating may cause the temperature
to rise further, hence, our reasoning adopts a conservative position.

\paragraph{The distribution of clouds in exoplanet atmospheres:}
The global distribution of clouds in exoplanet atmospheres depends on
system parameters like orbital distance, rotational period, mass,
radius, global temperature (represented in Fig.\ref{diversity}) and
stellar spectral type as these determine the local temperature
distribution inside a planet's atmosphere. Several of the known
exoplanets are tidally locked, hence, the same side of the planet
always receives the highest radiation flux from the host star and the
other is in permanent darkness (like the Moon). Giant gas planets like
HD\,189733b and HD\,209458b with an orbital distance of $\sim0.03$AU
and $\sim 0.05$AU, respectively, have a cloud cover that extends
across the whole globe
\cite{2016A&A...594A..48L,2018A&A...615A..97L}. Moving a little
further towards the host star increases the dayside temperature of
planets like WASP-18b (orbital distance of $\sim0.02$AU) such that
clouds only form on the nightside and in some terminator regions. The
dayside remains largely cloud free
\cite{2019arXiv190108640H}. All known brown dwarfs that do not orbit a
white dwarf will have a global cloud coverage comparable to giant gas
planets like HD\,189733b and HD\,209458b. Such a cloud coverage can,
however, be inhomogeneous showing holes, vortexes or banded structures
depending on the atmosphere dynamic which is affected by the brown
dwarf rotation rather than by an external radiation source
\cite{2014ApJ...788L...6Z}.  Recent observations have compared the
banded cloud structure of a brown dwarf with that observed on Jupiter
and Neptune in the Solar System \cite{2017Sci...357..683A}.

\section{Conclusion:  Which extrasolar planetary objects are favorable for developing a  extrasolar Global Electric Circuits?}  Every extrasolar planetary object, i.e. exoplanets and brown dwarfs, that harbors an atmosphere can establish a Global Electric Circuit (GEC) if lightning is possible. Lightning will occur where dynamic clouds form. This depends on the planet's and brown dwarf's system parameters which change as the planet  / brown dwarf evolves. Given the diversity of exoplanets, detailed studies are required for individual objects but  three general scenarios can be considered in order to decide  which objects may be favorable for  extrasolar Global Electric Circuits:
\begin{itemize}
\item[ A)] {\it Full cloud coverage:} If the exoplanet is fully
  covered in clouds, fair-weather regions through which a leaking
  current equilibrates large-scale electric potential differences will
  be difficult to establish but lightning may occur easily. A global
  circuit system might fail to establish. However, smaller circuit
  systems may develop on local scales. Giant gas planets that are not
  too close to their host star (e.g. HD\,189733b and HD\,209458b) are
  fully covered in clouds which extent over several vertical pressure
  scale heights.  Repeated observations of the super-Earth GJ1214b
  \cite{2014Natur.505...69K} suggest that a geometrically thick cloud
  coverage is present here, too. Also most brown dwarfs may fail to
  establish a eGCE because they are fully covered in clouds unless
  they are part of a highly-irradiating white dwarf binary system.
\item[ B)] {\it Intermittent cloud coverage:} Earth is an example for
  intermittent cloud coverage where cloudy and fair-weather regions
  characterise the atmosphere. Observations point to variable cloud
  properties (like height, particle size) on brown dwarfs
  \cite{2017Sci...357..683A}, but it is impossible to draw conclusions
  about implications for the presence of fair-weather regions. No such
  measurements can be obtained for exoplanets with the present
  technology.
\item[ C)] {\it Partial cloud coverage:} Ultra-hot giant gas planets
  (e.g. WASP-18b, HAT-P-7b) form clouds predominantly on their night
  side but no clouds on the dayside. In addition, the higher local
  temperature does increase the dayside's atmospheres local
  conductivity compared to the nightside. Brown dwarfs that are part
  of a white-dwarf binary system may fall into this category
  (e.g. WD0137-349B \cite{2017MNRAS.471.1728L}), too. Both classes of
  objects will, hence, develop a fair-weather dayside and a cloudy,
  lightning producing nightside. Exoplanet objects that develop a
  partial cloud coverage may therefore develop a global electric
  circuit over larger scales than known for Earth.
\end{itemize}

\ack
The funding from the European commission is acknowledged under which part of this research was conducted. We highlight financial support of the European Union under the FP7 by an ERC starting grant number 257431.

\section*{References}

\bibliographystyle{iopart-num}
\bibliography{bib.bib}

\end{document}